\documentclass[10pt,twocolumn,letterpaper]{article}

\usepackage{cvpr}              

\usepackage{booktabs}
\usepackage[dvipsnames]{xcolor}
\usepackage[table]{xcolor}
\usepackage{colortbl}
\usepackage{rotating}
\usepackage{makecell}
\usepackage{graphicx}
\usepackage{subcaption}
\usepackage{wrapfig}
\usepackage{tabularx}
\newcolumntype{Y}{>{\centering\arraybackslash}X}
\usepackage{array}
\usepackage{amsmath}
\definecolor{cvprblue}{rgb}{0.21,0.49,0.74}
\usepackage[pagebackref,breaklinks,colorlinks,allcolors=cvprblue]{hyperref}

\def\paperID{27925} 
\def\confName{CVPR}
\def\confYear{2026}

\title{Detecting and Mitigating Insertion Hallucination in Video-to-Audio Generation}

\author{
\textbf{Liyang Chen}$^{1,3}$ \quad
\textbf{Hongkai Chen}$^{3}$ \quad
\textbf{Yujun Cai}$^{2}$ \\
\textbf{Sifan Li}$^{3,4}$ \quad
\textbf{Qingwen Ye}$^{3}$ \quad
\textbf{Yiwei Wang}$^{4}$ \\[6pt]
$^{1}$University of California, Los Angeles \quad
$^{2}$The University of Queensland \\
$^{3}$vivo Mobile Communication Co., Ltd. \quad
$^{4}$University of California, Merced
}

\begin{document}

\maketitle

\begin{abstract}
Video-to-Audio generation has made remarkable strides in automatically synthesizing sound for video. However, existing evaluation metrics, which focus on semantic and temporal alignment, overlook a critical failure mode: models often generate acoustic events, particularly speech and music, that have no corresponding visual source. We term this phenomenon Insertion Hallucination and identify it as a systemic risk driven by dataset biases, such as the prevalence of off-screen sounds, that remains completely undetected by current metrics. To address this challenge, we first develop a systematic evaluation framework that employs a majority-voting ensemble of multiple audio event detectors. We also introduce two novel metrics to quantify the prevalence and severity of this issue: IH@vid (the fraction of videos with hallucinations) and IH@dur (the fraction of hallucinated duration). Building on this, we introduce HALCON to mitigate IH. HALCON follows a three-stage procedure: it first generates initial audio to expose hallucinated segments, then identifies and masks the corresponding unreliable video features, and finally regenerates the audio using the corrected conditioning. Experiments on several mainstream V2A benchmarks first reveal that state-of-the-art models suffer from severe IH. In contrast, our HALCON method reduces both the prevalence and duration of hallucinations by over 50\% on average, without degrading, and in some cases even improving, conventional metrics for audio quality and temporal synchronization. Our work is the first to formally define, systematically measure, and effectively mitigate Insertion Hallucination, paving the way for more reliable and faithful V2A models.
\end{abstract}

\section{Introduction}

Sound design is essential for realism and immersion in films, games, animations, and other multimedia content, as audio provides temporal, spatial, and emotional cues that silent visuals alone cannot convey. Traditional Foley production, which involves manually recording, editing, and mixing sound effects, is highly specialized and difficult to scale, motivating the development of automatic sound generation systems. Recent progress in Video-to-Audio (V2A) generation has shown promise in this direction: models such as MMAudio~\citep{cheng2025mmaudio} and ThinkSound~\citep{liu2025thinksound} learn semantic and temporal audio-video alignment from large-scale paired datasets and achieve impressive performance. Their evaluations typically rely on metrics like FD-VGG~\citep{fid,vggish}, ISC~\citep{isc}, and DeSync~\citep{desync}, which assess semantic similarity and temporal synchronization and have driven substantial advances in generating the right sounds at the right moments.

\begin{figure}[t]
    \centering
    
    \begin{subfigure}{1\linewidth} 
        \centering
        \includegraphics[width=\linewidth]{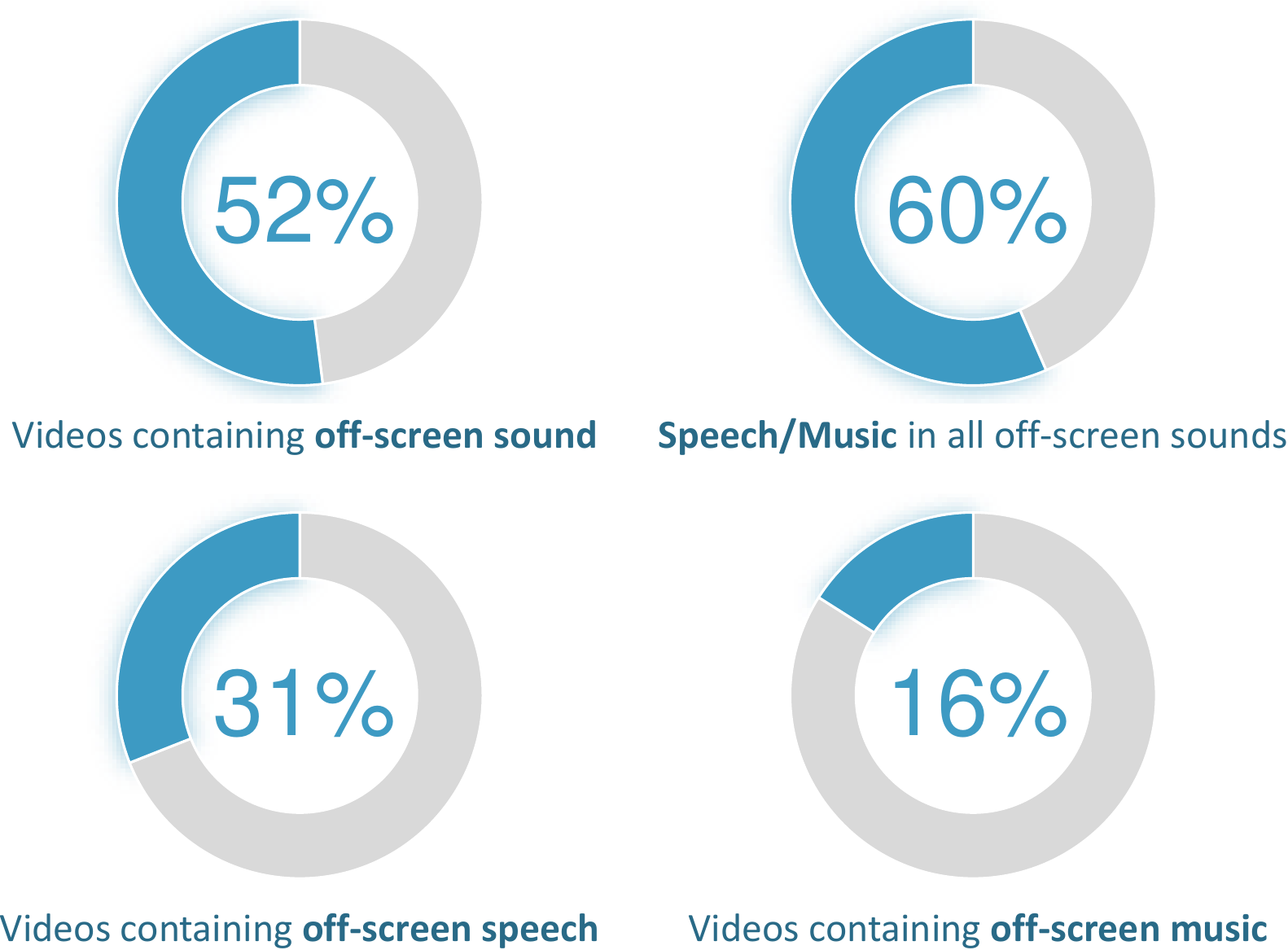}
        \caption{Overview of off-screen sound statistics, including its prevalence and the proportions of speech and music within off-screen events.}
        \label{fig:vgg_stats_1}
    \end{subfigure}
    \vspace{6pt}
    \begin{subfigure}{1\linewidth} 
        \centering
        \includegraphics[width=\linewidth]{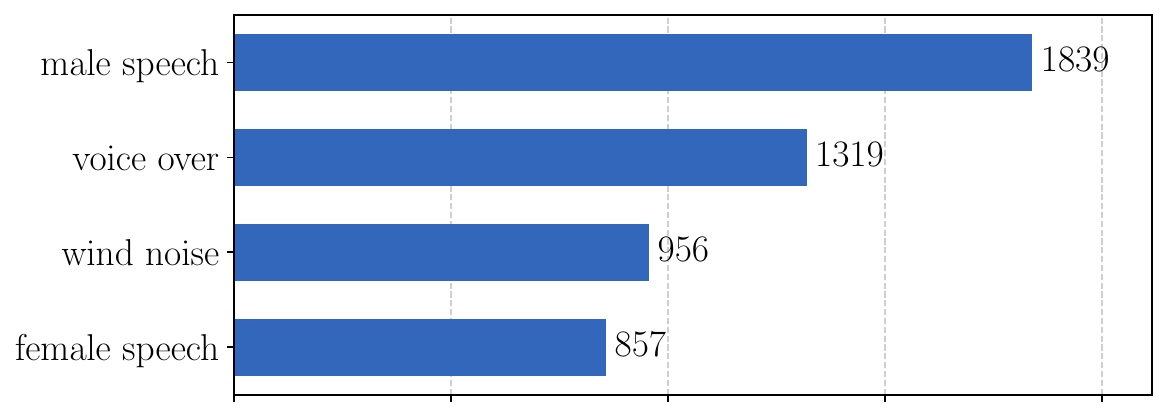}
        \caption{Category-level distribution of common off-screen sounds, showing that various forms of speech dominate.}
        \label{fig:vgg_stats_2}
    \end{subfigure}

    \caption{Statistics of off-screen sound in the VGGSound dataset, illustrating its high prevalence and the strong dominance of speech and music within off-screen events.}
    \label{fig:vgg_stats}
\end{figure}

\begin{figure*}[h]
    \centering
    \includegraphics[width=1\linewidth]{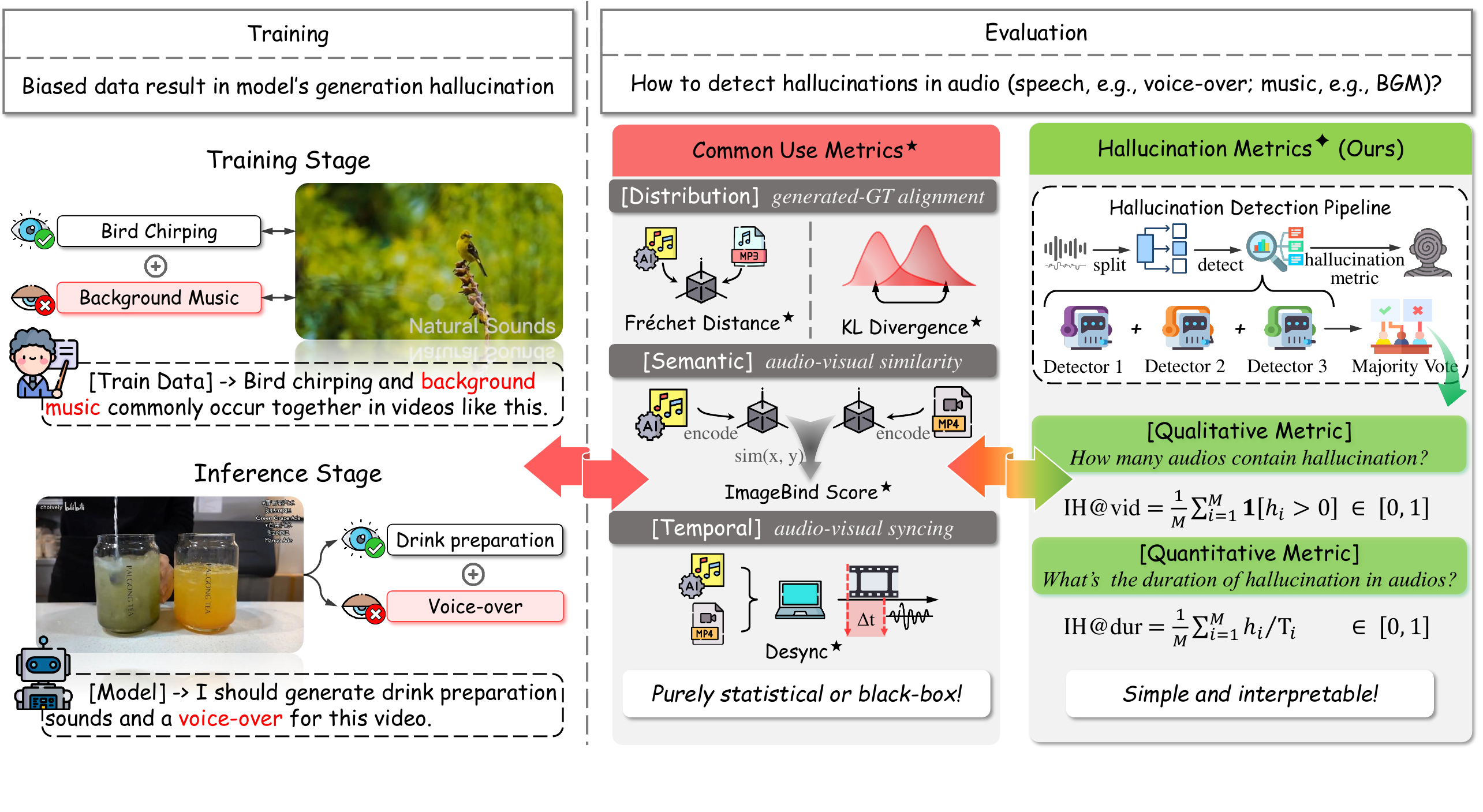}
    \caption{Example of Insertion Hallucination (IH) in video-to-audio generation. 
    Training data often include off-screen speech or music, which biases models to reproduce them. 
    As a result, during inference the model may generate speech or music even when the video only depicts other events.}
    \label{fig:ih_example}
\end{figure*}

However, we observe that existing V2A models frequently generate sounds that do not correspond to the visual content---for example, producing melodic music for a sanding scene or synthesizing human voices in a vacuum-cleaner video. As illustrated in Figure~\ref{fig:ih_example}, such errors often stem from biased audio–visual co-occurrence patterns in training data, which encourage models to associate unrelated events. Yet current evaluation metrics fail to capture these mistakes. In practice, around 50\% of VGGSound samples contain off-screen sounds, most of which fall into speech or music categories (Figure~\ref{fig:vgg_stats}; \citet{vggsounder}). This bias predisposes models to hallucinate speech or music when visual cues are weak, while remaining entirely undetected by existing metrics, leading to misleading assessments of model reliability. We refer to this overlooked phenomenon as \textit{Insertion Hallucination (IH)}, which denotes the generation of sound events that have no visual counterpart. Through empirical studies, we find that models such as ThinkSound~\citep{liu2025thinksound} and MMAudio~\citep{cheng2025mmaudio} frequently exhibit IH on mainstream datasets including VGGSound~\citep{vggsound} and Kling-Audio-Eval~\citep{klingfoley}. These findings indicate that IH is a systematic and pervasive risk in V2A generation, yet it remains largely undetected by existing evaluation metrics and unaddressed in prior work.

To systematically investigate IH, we develop an end-to-end evaluation framework. 
We begin with an automatic detection pipeline that identifies hallucinated segments 
by integrating three audio event detectors: inaSpeechSegmenter~\citep{inaseg}, 
YAMNet~\citep{yamnet}, and PANNs~\citep{panns}, and fusing their outputs through 
majority voting. The pipeline is then validated on a human-annotated set to assess 
its accuracy. Finally, we introduce two metrics, IH@vid (the fraction of videos 
containing hallucination) and IH@dur (the fraction of hallucinated duration), 
to quantify the frequency and severity of IH in V2A models.

Building on this, we propose a novel inference-time conditioning method called HALCON. HALCON does not require retraining the model and instead operates in three stages. In Stage 1, the model generates an initial audio prediction from the input video. In Stage 2, we detect hallucinated segments in the generated audio and correct them by masking the corresponding video features with empty features. In Stage 3, the model regenerates the audio using the corrected video features. This procedure encourages the model to rely on contextual or label information rather than unreliable visual cues, preventing it from degenerating into speech or music hallucinations. Experiments show that HALCON significantly reduces IH@vid and IH@dur while preserving standard metrics.

Our main contributions are summarized as follows:
\begin{itemize}
\item We are the first to define Insertion Hallucination (IH) in audio generation, revealing realism as a critical risk dimension that is completely overlooked by existing evaluation metrics.
\item We build an IH evaluation framework combining multi-detector voting and human verification, and propose two metrics (IH@vid and IH@dur) to quantify models’ hallucination tendency.
\item We propose HALCON, a HAlucination-enhanced CONditioning method that significantly reduces IH while maintaining conventional metrics, and demonstrate its effectiveness on multiple V2A benchmarks.
\end{itemize}

\section{Related Work}

\subsection{Video-to-Audio Generation}
Early work on Video-to-Audio generation was dominated by Generative Adversarial Networks \citep{chatterjee2020sound2sight,pascual2017segan,ferreira2022generation}, but recent advances have shifted toward models that produce higher-quality audio with stronger audio-visual alignment. A key direction is improving representation alignment: Diff-Foley \citep{luo2023diff} employs contrastive audio-visual pretraining to learn temporally and semantically aligned features that guide a latent diffusion model, achieving substantial gains in synchronization and relevance.

With the rise of more capable generative models, research has expanded toward controllability and practicality. FoleyCrafter \citep{zhang2024foleycrafter} adapts a pre-trained text-to-audio model with semantic and temporal controllers, enabling prompt-based control with precise alignment. Data expansion is another path: MMAudio \citep{cheng2025mmaudio} unifies video-audio and large-scale text-audio data for richer semantics, while MultiFoley \citep{chen2025video} conditions on both text and audio for flexible user guidance and high-fidelity synchronization.

The frontier is now moving beyond direct mapping to incorporate reasoning. ThinkSound \citep{liu2025thinksound} introduces a Chain-of-Thought framework in which a multimodal large language model produces interpretable reasoning steps that guide audio generation, transforming the task into a cognitively driven process.

Nevertheless, evaluation remains centered on semantic relevance and synchronization, neglecting whether generated sounds should appear in the video at all. Current metrics cannot capture hallucinations such as spurious speech or music, leaving a critical gap that our work aims to fill.

\subsection{Hallucination in Large Language Models}
Hallucination is a core challenge for large-scale AI models, with related research expanding from Large Language Models to the multimodal domain. In Large Language Models, researchers mitigate insertion hallucination by introducing external knowledge bases \citep{lee2022factuality} or enhancing internal consistency \citep{mundler2023self}. This issue manifests in Vision-Language Models as object hallucination, where a model describes non-existent objects in an image. The academic community has established dedicated evaluation benchmarks such as POPE \citep{Li-hallucination-2023} and proposed solutions such as Object-Aware Preference Optimization \citep{chen2024multi,compagnoni2025mitigating}. Recently, the evaluation of hallucination has extended to the audio-visual domain; for instance, AVHBench \citep{sung2024avhbench} designs cross-modal understanding tasks to assess whether a model exhibits audio-driven or video-driven hallucinations. However, while these studies prove that hallucination is a common risk in multimodal models, the evaluation endpoint of all existing work, whether for language, vision, or audio-visuals, is exclusively focused on whether the generated textual output contains hallucinations. The phenomenon of the audio itself being the subject of hallucination, such as a Video-to-Audio model generating sound that contradicts the visual scene, remains an unexplored research gap.

\subsection{Off-screen Sound Generation}
A notable recent trend in the V2A field is the research on generating off-screen sound. Many researchers have observed that existing video datasets commonly contain off-screen audio events, and they aim to make models learn and align with this characteristic to generate more complete and immersive holistic soundscapes. For instance, VinTAGe \citep{kushwaha2025vintage} leverages additional information such as text to assist in generating off-screen sounds, while Action2Sound \citep{chen2024action2sound} independently models off-screen ambient audio by separating it from foreground sounds. The importance of this trend is also reflected in the evolution of evaluation methods: VGGSounder \citep{zverev2025vggsounder} was the first to introduce an off-screen sound dimension into its evaluation framework. By comparing model performance with and without visual cues, it revealed a common "over-reliance on vision" bias in existing models, thereby emphasizing the importance of independent audio understanding capabilities.

However, we argue that pursuing alignment with off-screen sounds poses a risk, as it may sacrifice the model's fidelity to the visual content and its generalization capabilities. In contrast, we advocate for the "What You See Is What You Get" principle. We believe that a model should first focus on generating faithful and reliable audio for visible visual content, as this is the fundamental basis for building controllable and trustworthy generative models.

\section{Methodology for Measuring Insertion Hallucination}
\label{sec:method}

\subsection{Problem Analysis and Definition}

Video-to-Audio generation models learn a conditional mapping $P(a|v)$ from visual input $v$ to audio output $a$ using paired training data. However, mainstream datasets contain a high prevalence of off-screen sounds, particularly speech and music, which introduces a systematic bias. When visual cues are weak or ambiguous, models often default to reproducing these frequent patterns rather than faithfully rendering scene-consistent audio.

We define this failure mode as \emph{Insertion Hallucination} (IH): the generation of structured acoustic events that have no corresponding source in the visual content. While IH could in principle include any spurious sound, we focus on speech and music for three reasons: (1) they are the most frequent off-screen sounds in mainstream corpora, with over half of VGGSound samples exhibiting this bias (Figure~\ref{fig:vgg_stats}); (2) they are perceptually salient events whose presence can strongly disrupt immersion; and (3) mature detection tools are available, enabling reliable identification.

Formally, given a video–label pair $(v, y)$, where $y$ specifies the ground-truth sound category, and an audio prediction $\hat a = G(v)$ from a model $G$, we define the hallucination indicator as:
\[
\text{is\_IH}(v, y, \hat{a}) =
\begin{cases}
1, & \text{if } y \notin \mathcal{Y}_{sm} \;\text{ and }\; D(\hat{a}) \neq \emptyset, \\
0, & \text{otherwise},
\end{cases}
\]
where $\mathcal{Y}_{sm}$ is the set of speech and music labels and $D(\hat{a})$ denotes detected hallucinated segments.

\subsection{Multi-Detector Ensemble Framework}

Detecting hallucinations reliably requires addressing the limitations of individual audio classifiers. To this end, we design a multi-detector ensemble that combines three complementary detectors: inaSpeechSegmenter \citep{inaseg}, YAMNet \citep{yamnet}, and PANNs \citep{panns}.  

Our pipeline consists of three stages:
\begin{enumerate}
    \item \textbf{Candidate Filtering.} Samples with ground-truth labels in $\mathcal{Y}_{sm}$ are excluded, ensuring that evaluation is limited to videos where speech and music are not expected. 
    \item \textbf{Multi-Detector Analysis.} Each detector independently identifies speech and music segments based on its model-specific decision boundary.  
    \item \textbf{Ensemble Fusion.} To combine the outputs of multiple detectors, we consider 
    three fusion strategies: (1) \textit{AND}, which marks a segment as hallucinated only if 
    all detectors agree; (2) \textit{OR}, which accepts any detector’s positive prediction; 
    and (3) \textit{Majority Vote (MV)}, which assigns a hallucination label when at least 
    half of the detectors vote positive. The MV rule is formally written as:
    \[
    D_{R}(\hat{a}) = \big\{ s \,\big|\, \sum_{k=1}^{K} \mathbf{1}[s \in D_k(\hat{a})] 
    \ge \lceil K/2 \rceil \big\}.
    \]
    A full comparison of these fusion strategies and the justification for choosing MV 
    are presented in Sec.~\ref{sec:validation}.

\end{enumerate}

This ensemble balances precision and recall while being robust to detector-specific biases. We validate its reliability against human annotations (Section~\ref{sec:experiments}).

\subsection{Evaluation Metrics}

To quantify hallucination behavior, we introduce two complementary metrics.  
Let $M$ denote the number of evaluated samples, $d_i$ the total hallucinated duration of sample $i$, and $T_i$ its total length.

\[
\mathrm{IH@vid} = \frac{1}{M} \sum_{i=1}^{M} \mathbf{1}[d_i > 0], 
\qquad
\mathrm{IH@dur} = \frac{1}{M} \sum_{i=1}^{M} \frac{d_i}{T_i}.
\]

IH@vid measures the proportion of audios that contain hallucination (prevalence), while IH@dur measures the proportion of hallucinated duration (severity).

\begin{figure*}[t]
    \centering
    \includegraphics[width=\linewidth]{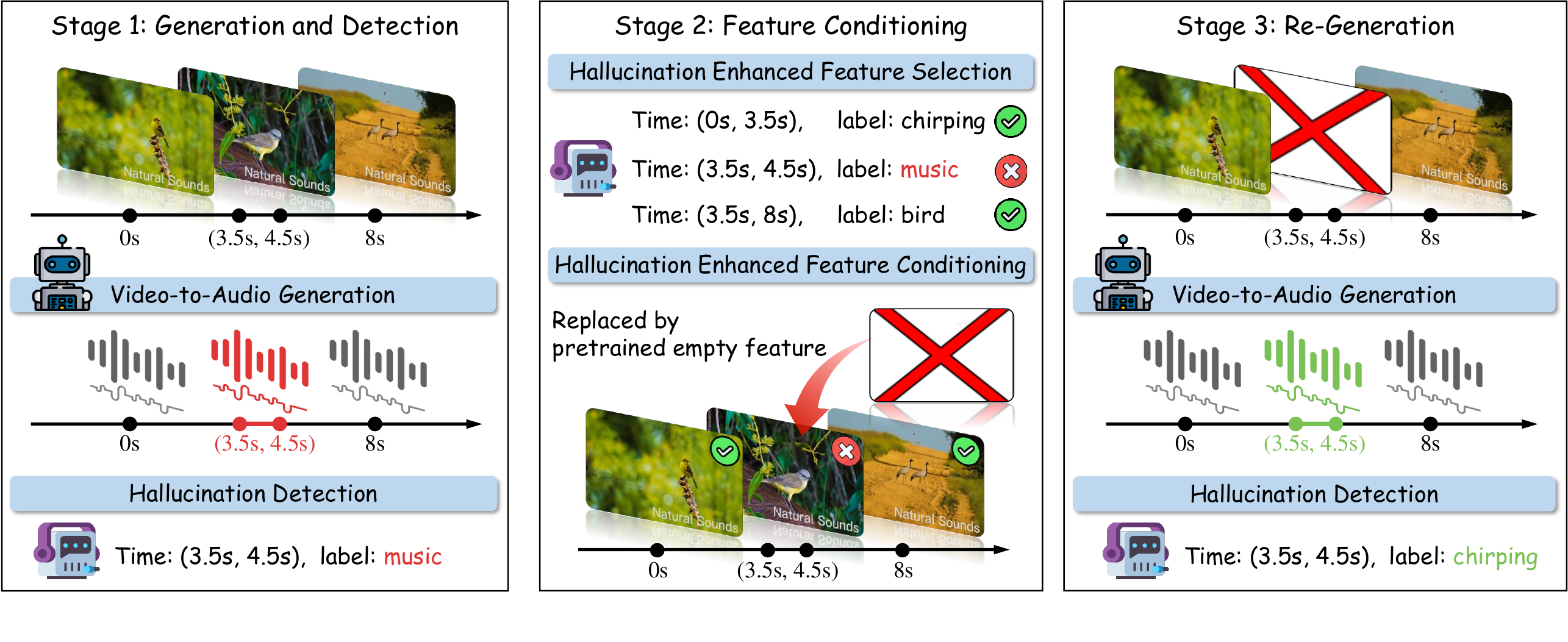}
    \caption{Overview of our proposed method HALCON.}
    \label{fig:pfc_overview}
\end{figure*}

\section{Hallucination-enhanced Conditioning}

Despite recent advances, state-of-the-art V2A models still exhibit Insertion Hallucination when visual inputs provide insufficient cues for reliable generation. This failure mode reflects a systematic reliance on strong dataset priors, in particular speech and music, whenever the visual signal is ambiguous. To address this issue, we propose HALCON, a HALlucination-enhanced CONditioning method that dynamically masks unreliable video features identified through hallucination detection.

\subsection{Method Motivation}

We observe that hallucinations arise most often when visual representations fail to provide discriminative guidance for audio generation. This suggests a feedback mechanism: if we can identify where hallucinations occur, we can infer where visual features are unreliable.  

The key insight is that V2A models exhibit predictable failure modes. When visual encoders produce ambiguous representations, for example due to visual similarity between acoustically different events, poor lighting, or out-of-distribution content, models fall back to generating high-frequency training patterns. The location of these hallucinations thus serves as a diagnostic signal for visual uncertainty. 

We propose HALCON, which exploits this signal through a three-stage process: first generate audio to identify problematic regions, then detect and correct the corresponding visual features, and finally regenerate the audio with the corrected features. By removing unreliable visual cues, we force the model to rely on more conservative generation strategies and stronger contextual information.

This approach is inspired by self-correction mechanisms in other domains~\citep{madaan2023self,shinn2023reflexion,huang2023large}, 
but uniquely leverages the temporal structure of audio-visual alignment for targeted feature intervention. An overview of the process is shown in Figure~\ref{fig:pfc_overview}.

\subsection{Algorithm Design}

HALCON operates in three inference stages:

\paragraph{Stage 1: Initial Generation.}
Given an input video $v$ with visual features $f_v$, the model generates an initial audio prediction 
$\hat{a} = G(v, f_v)$.

\paragraph{Stage 2: Hallucination Detection and Feature Correction.}
We apply our hallucination detector $D(\cdot)$ to obtain a set of hallucination intervals 
$\mathcal{H} = D(\hat{a})$, where each $\tau = [s,e] \in \mathcal{H}$ denotes a time span 
predicted as a speech or music hallucination. 
We then construct corrected visual features $f'_v$ by replacing the features at hallucinated 
timestamps with an empty feature $\emptyset_v$, a neutral representation learned during the 
model's pretraining to encode the absence of visual information. This empty feature behaves 
as an in-distribution “no-vision” token, preventing the model from being influenced by 
misleading visual cues:

\[
f'_v(t) =
\begin{cases}
\emptyset_v, & t \in \bigcup_{\tau \in \mathcal{H}} \tau, \\
f_v(t), & \text{otherwise}.
\end{cases}
\]

\paragraph{Stage 3: Re-Generation.}
The corrected features $f'_v$ are fed back into the same model to obtain a revised audio output 
$\hat{a}' = G(v, f'_v)$.

This three-stage procedure uses hallucination locations as uncertainty indicators and intervenes 
only where necessary. By removing unreliable visual cues, the model is encouraged to rely on 
contextual or label information, thereby reducing IH while preserving semantic accuracy and 
temporal synchronization elsewhere.

\section{Experiments}
\label{sec:experiments}

This section presents a comprehensive empirical validation of our proposed framework. We first validate our IH detection pipeline on a human-annotated dataset (Section~\ref{sec:validation}). We then apply it to assess the prevalence of Insertion Hallucination (IH) in state-of-the-art models and evaluate HALCON (Section~\ref{sec:assessment}). Finally, we analyze HALCON's core components via an ablation study (Section~\ref{sec:ablation}) and compare it against alternative correction methods (Section~\ref{sec:comparison_alternatives}).

\begin{figure*}[h]
    \centering
    
    \begin{subfigure}[c]{0.48\linewidth} 
        \centering
        \begin{tabularx}{\linewidth}{lYYYY}
        \toprule
        \textbf{Method} & \textbf{TP (\%)} & \textbf{FP (\%)} & \textbf{FN (\%)} & \textbf{IoU (\%)} \\
        \midrule
        AND & 13.66 & 1.28 & 11.39 & 51.90 \\
        OR  & 21.71 & 7.49 & 3.34  & 66.72 \\
        MV  & 18.46 & 3.14 & 6.59  & 65.47 \\
        \bottomrule
        \end{tabularx}
        \caption{Percentage-based comparison of TP, FP, FN, and IoU.}
        \label{tab:ih_validation_percentage}
    \end{subfigure}
    \begin{subfigure}[c]{0.50\linewidth} 
        \centering
        \includegraphics[width=\linewidth]{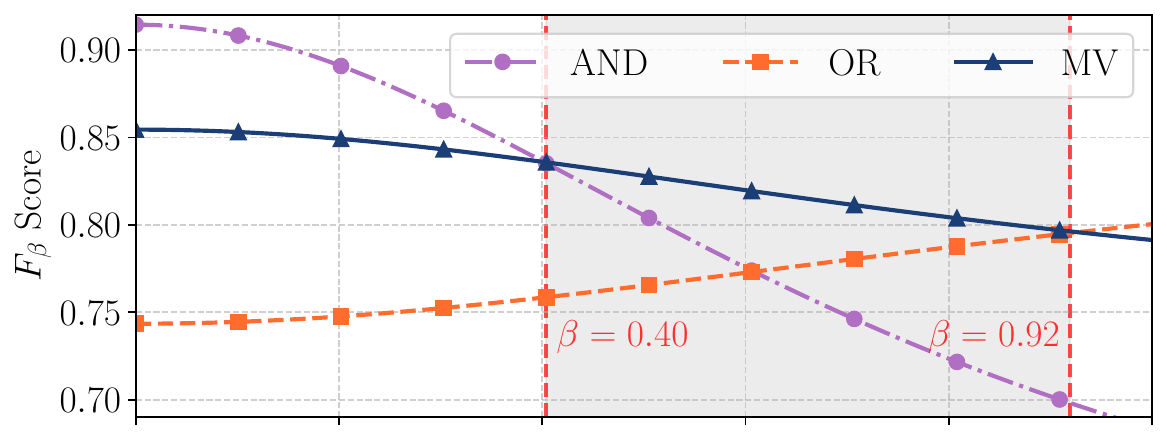}
        \caption{$F_{\beta}$ curves across varying $\beta$ values.}
        \label{fig:f_beta_curve}
    \end{subfigure}

    \caption{Combined evaluation of fusion strategies. The table reports percentage-based metrics, while the curves illustrate the behavior of $F_{\beta}$ under different precision--recall weightings.}
    \label{fig:combined_performance} 
\end{figure*}

\subsection{Experimental Setup}

We validated our IH detection pipeline using three audio event detectors, inaSpeechSegmenter, YAMNet, and PANNs, on the human-annotated set described in Appendix, measuring Precision, Recall, $F_{\beta}$-score, and IoU. We then applied the validated metrics to evaluate two representative V2A systems: MMAudio \citep{cheng2025mmaudio}, a multimodal framework with synchronization and flow-matching generation, and ThinkSound \citep{liu2025thinksound}, a reasoning-based model that uses Chain-of-Thought for visual and temporal modeling. Experiments were conducted on three benchmarks: Kling-Audio-Eval \citep{klingfoley} (20k clips, 1.9k classes, stricter filtering of off-screen sounds), VGGSound \citep{vggsound} (200k clips, 310 classes, known off-screen bias), and AVE \citep{tian2018ave} (4,143 clips, 28 classes, frame-level annotations). Evaluation covered hallucination specific metrics (IH@vid, IH@dur), distributional metrics (FD\textsubscript{PANNs}, KL\textsubscript{PANNs}), semantic/quality metrics (ISC, IB-score), and temporal alignment (DeSync).

\subsection{Validation of IH Metrics}
\label{sec:validation}

We validated the reliability of our IH detection pipeline on a dedicated human annotated dataset, where clips were manually labeled for the presence of speech and music hallucinations (see Appendix). However, human annotators are generally insensitive to fine-grained acoustic boundaries and often overlook short pauses on the order of tens or even hundreds of milliseconds. As a result, relying solely on recall oriented evaluation can be misleading. A more reliable criterion is ensuring that each detected segment corresponds to a genuine human verified event, which calls for placing stronger emphasis on \textit{precision}.

To quantify this emphasis, we adopt the generalized $F_{\beta}$-score:
\[
F_{\beta} = (1 + \beta^2) \cdot \frac{\text{Precision} \cdot \text{Recall}}
{\beta^2 \cdot \text{Precision} + \text{Recall}}.
\]
Precision oriented evaluation corresponds to $\beta < 1$, though choosing $\beta$ too small would overly penalize recall and suppress meaningful detections. As shown in Fig.~\ref{fig:f_beta_curve}, the \textbf{Majority Vote (MV)} fusion strategy consistently achieves the highest $F_{\beta}$ scores across the range $\beta \in [0.40, 0.92]$, outperforming both AND and OR. This interval also aligns with our goal of favoring precision without overcommitting to it, as it emphasizes precision while still preserving adequate recall.

The strong and stable performance of MV across this precision leaning interval indicates that it offers the most reliable trade-off between false positives and false negatives. We therefore adopt \textbf{MV} as the fusion strategy for all subsequent experiments.

\begin{table*}[t]
\centering
\setlength{\tabcolsep}{10pt} 
\caption{Results on Kling-Audio-Eval, VGGSound, and AVE, showing that HALCON consistently reduces hallucinations without degrading quality or synchronization.}
\label{tab:merged-results}
\newcommand{\mycolsep}{\hspace{0.2em}}

\definecolor{CellColor}{HTML}{FBF1B7}
\newcommand{\hlcell}[1]{\cellcolor{CellColor!50} #1}

\definecolor{TextColor}{HTML}{228B22}
\newcommand{\celldown}[1]{\textcolor{TextColor}{$\downarrow$ #1}}

\begin{tabular}{ l *{7}{c} }
\toprule
& \multicolumn{5}{c}{\textbf{Common Use Metrics}} & \multicolumn{2}{c}{\textbf{Hallucination Metrics}} \\
\cmidrule(r){2-6} \cmidrule(l){7-8}
%
\textbf{Method} & \textbf{FD $\downarrow$} & \textbf{KL $\downarrow$} & \textbf{ISC $\uparrow$} & \textbf{IB $\uparrow$} & \textbf{DeSync $\downarrow$} & \textbf{IH@vid $\downarrow$} & \textbf{IH@dur $\downarrow$} \\
\midrule
\multicolumn{8}{c}{Kling-Audio-Eval} \\  
\cmidrule{1-8}
GT            & --    & --    & --    & --    & --    & 5.2 & 1.4 \\
MMAudio \citep{cheng2025mmaudio}       & 10.48 & 2.50 & 8.34 & 0.34 & 0.61 & 12.9 & 4.6 \\
+ HALCON         & 10.96 & 2.46 & 8.23 & 0.34 & 0.61 & \hlcell{6.1 \mycolsep \celldown{52.4\%}} & \hlcell{2.5 \mycolsep \celldown{46.2\%}} \\
ThinkSound \citep{liu2025thinksound}    & 12.48 & 2.76 & 5.49 & 0.21 & 0.74 & 24.3 & 14.7 \\
+ HALCON         & 12.35 & 2.53 & 5.43 & 0.22 & 0.71 & \hlcell{9.2 \mycolsep \celldown{62.0\%}} & \hlcell{5.2 \mycolsep \celldown{64.7\%}} \\

\midrule
\multicolumn{8}{c}{VGGSound} \\  
\cmidrule{1-8}
GT            & -- & -- & -- & -- & -- & 11.0 & 2.6 \\
MMAudio \citep{cheng2025mmaudio}       & 6.87 & 1.81 & 7.01 & 0.34 & 0.60 & 16.3 & 6.1 \\
+ HALCON         & 6.49 & 1.78 & 7.13 & 0.34 & 0.59 & \hlcell{8.9 \mycolsep \celldown{45.4\%}} & \hlcell{5.5 \mycolsep \celldown{10.2\%}} \\
ThinkSound \citep{liu2025thinksound}    & 6.67 & 2.02 & 5.73 & 0.23 & 0.72 & 13.0 & 5.2 \\
+ HALCON         & 6.57 & 1.95 & 5.80 & 0.22 & 0.72 & \hlcell{6.3 \mycolsep \celldown{51.9\%}} & \hlcell{3.9 \mycolsep \celldown{25.4\%}} \\

\midrule
\multicolumn{8}{c}{AVE} \\  
\cmidrule{1-8}
GT            & -- & -- & -- & -- & -- & 15.3 & 1.6 \\
MMAudio \citep{cheng2025mmaudio}       & 3.21 & 1.47 & 6.49 & 0.38 & 0.55 & 13.0 & 3.1 \\
+ HALCON         & 3.24 & 1.46 & 6.52 & 0.38 & 0.56 & \hlcell{6.1 \mycolsep \celldown{53.5\%}} & \hlcell{1.8 \mycolsep \celldown{40.7\%}} \\
ThinkSound \citep{liu2025thinksound}    & 8.23 & 1.92 & 5.43 & 0.25 & 0.72 & 19.1 & 7.4 \\
+ HALCON         & 7.39 & 1.95 & 5.36 & 0.25 & 0.72 & \hlcell{10.2 \mycolsep \celldown{46.4\%}} & \hlcell{3.0 \mycolsep \celldown{59.2\%}} \\
\bottomrule
\end{tabular}
\end{table*}

\subsection{Insertion Hallucination Assessment}
\label{sec:assessment}

We next apply our validated IH metrics to state-of-the-art V2A models and to evaluate the effectiveness of our proposed HALCON method.

\textbf{Baseline models exhibit systematic hallucination.}  
Table~\ref{tab:merged-results} reports results across Kling-Audio-Eval, VGGSound, and AVE. Both MMAudio and ThinkSound generate hallucinations in a substantial portion of videos (IH@vid 12--24\%), with spurious speech or music often occupying 4--15\% of the total duration. These findings establish that IH is not a rare anomaly but a widespread failure pattern in current V2A systems. The Ground-truth (GT) row, obtained by running our proposed IH detection pipeline on the dataset's reference audio, shows small non-zero IH values that reflect unavoidable dataset biases such as residual off-screen sounds or loosely aligned labels.

\textbf{HALCON substantially reduces hallucination.}
Across all benchmarks, HALCON consistently lowers hallucination rates. The effect is strongest on the more diverse Kling-Audio-Eval and AVE datasets, where IH@vid and IH@dur drop by 40 to 65\%. On VGGSound, the main training-domain dataset, HALCON still reduces hallucination frequency by 45 to 52\%, but the reduction in duration is smaller (10 to 25\%). This indicates that models overfit to training biases, making in-domain hallucinations harder to suppress, and highlights HALCON’s strength in improving generalization to out-of-domain data.

\textbf{Conventional metrics remain robust, showing no systematic degradation.}
Crucially, this targeted reduction in hallucination does not come at the cost of overall generation quality. As shown in the $\Delta$ rows, conventional metrics such as FD, KL, and DeSync exhibit only minor fluctuations, with most changes falling below 3\%. We even observe several instances of notable improvement, such as a 10.3\% enhancement in FD and an 8.4\% gain in KL for ThinkSound, suggesting that removing misleading visual features can sometimes help the model produce higher-quality audio. This stability confirms that HALCON is a non-destructive method that precisely targets unwanted content without degrading audio quality, diversity, or temporal alignment.

\subsection{Efficient Inference for HALCON}
\label{sec:efficient_inference}

The standard HALCON pipeline consists of three stages: an initial audio generation in Stage~1, 
a fast hallucination detection and feature correction in Stage~2, and a full regeneration in 
Stage~3. This three-stage generation design introduces substantial inference overhead, with Stage~1 being 
the primary computational bottleneck.

We therefore examine whether Stage~1 truly requires a high-fidelity generation. Our 
hypothesis is that a lower-fidelity audio---obtained with fewer sampling steps---may still 
reveal hallucination artifacts clearly enough for effective detection.

To test this, we vary the number of sampling steps $N$ used in Stage~1 on a subset of 
Kling-Audio-Eval, while keeping the Stage~3 generation fixed at 24 steps. After an 
early-terminated Stage~1, we immediately perform detection, correction, and a full 
24-step regeneration.

\begin{figure}[h]
  \centering
  \includegraphics[width=1\linewidth]{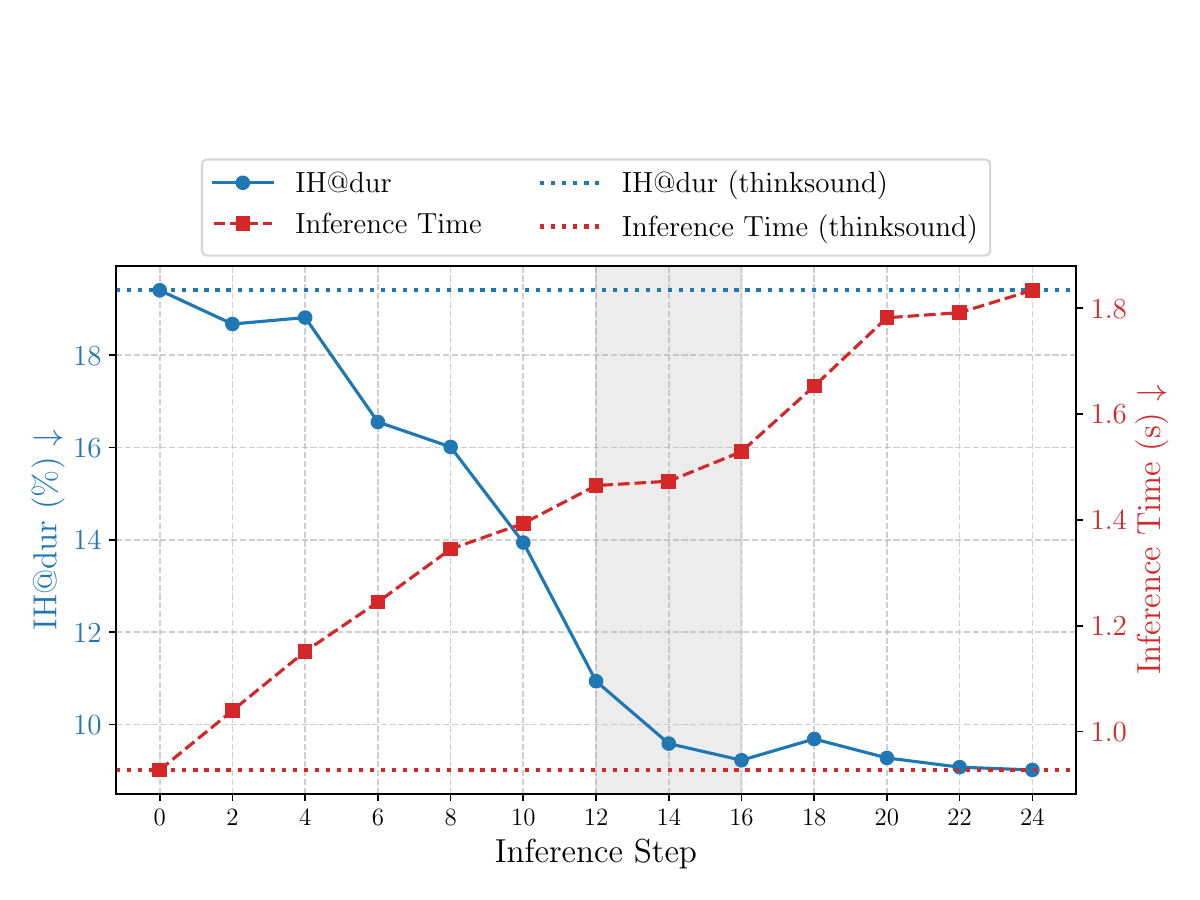}
  \caption{IH@dur and inference time versus Stage~1 sampling steps, highlighting the trade-off.}
  \label{fig:lightweight_inference}
\end{figure}

Figure~\ref{fig:lightweight_inference} reports the results: the blue curve shows the final 
IH@dur, and the red curve shows total inference time. Two observations emerge. First, 
small $N$ leads to low-fidelity audio that harms detection and yields high IH@dur, but 
performance rapidly improves and stabilizes as $N$ increases. Second, inference time grows 
almost linearly with $N$, with the full three-stage setting exceeding 1.8s.

The gray region highlights a favorable performance--efficiency trade-off. These findings 
show that HALCON's detection stage is robust to generation fidelity: using only a fraction of 
the sampling steps in Stage~1 achieves nearly the same hallucination suppression while 
substantially reducing end-to-end inference time. This greatly improves the practicality of 
deploying HALCON in real systems.

\subsection{Ablation of Feature Correction Strategies}
\label{sec:ablation}

To validate the effectiveness of our feature correction strategy, we conduct an ablation study to show that precisely targeting and correcting ``problematic'' visual features is essential. We compare our method (+HALCON) with two non-precise replacement strategies: random replacement (+Random) and complement replacement (+$\sim$HALCON), the latter modifying the non-hallucinated segments identified by our detector.

\begin{figure}[h]
  \centering
  \includegraphics[width=1\linewidth]{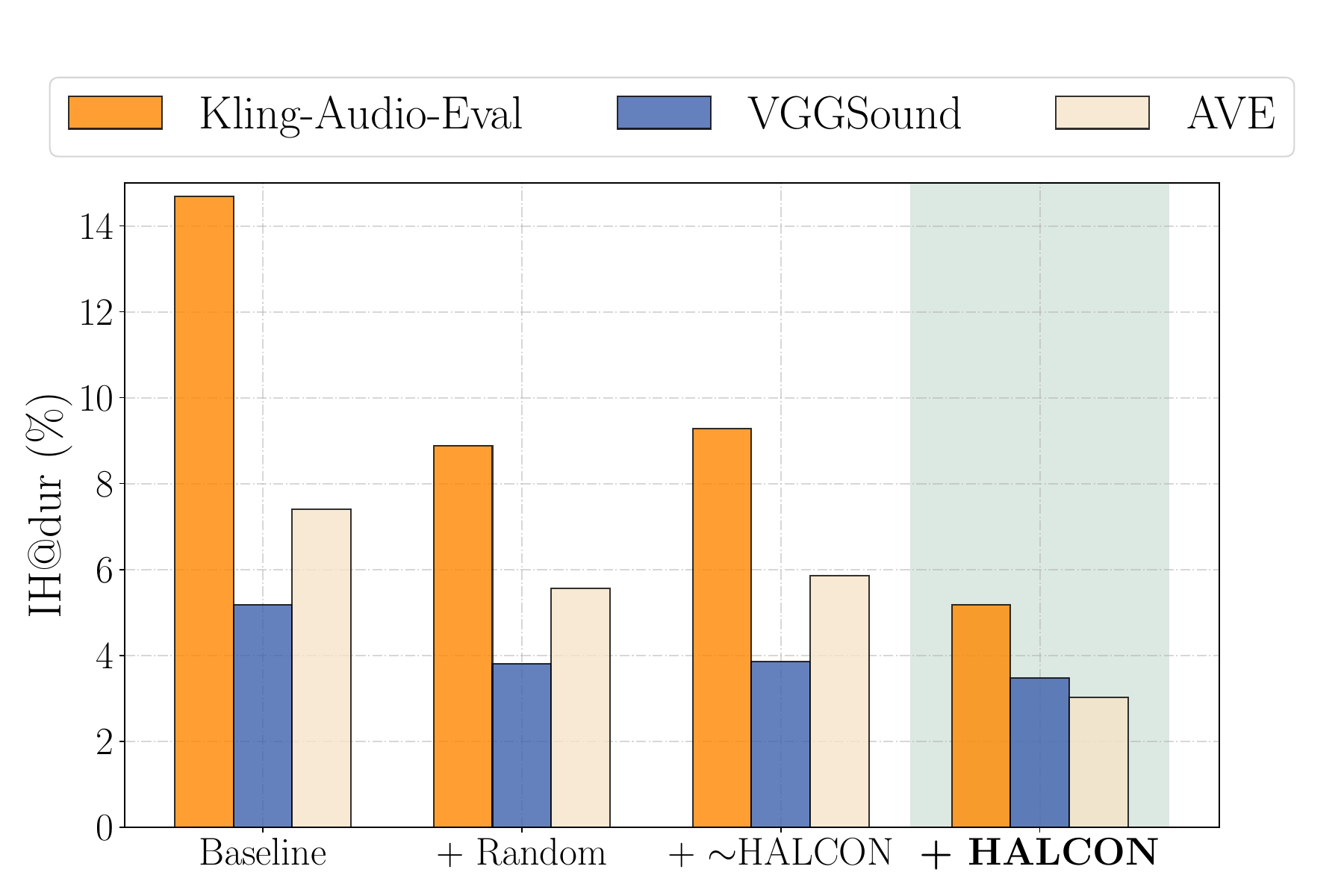}
  \caption{Ablation of feature correction strategies. HALCON yields the lowest IH@dur.}
  \label{fig:pfc_ablation}
\end{figure}

The results are shown in Figure~\ref{fig:pfc_ablation}. We first observe that all feature replacement strategies reduce IH@dur to some extent across datasets, suggesting that disrupting visual cues in general can weaken part of the hallucination-inducing signal. However, the improvements from non-precise strategies are limited and inconsistent. Notably, the complement replacement (+$\sim$HALCON) strategy even underperforms random replacement on the AVE dataset, indicating that altering non-problematic regions is not only unhelpful but may actively degrade model behavior. This confirms that the segments identified by our detector are indeed the critical ones that require correction.

In contrast, HALCON consistently achieves the lowest IH@dur on all datasets, with average performance substantially surpassing all alternatives. These findings demonstrate that HALCON’s effectiveness arises not merely from feature replacement, but from its ability to \emph{precisely} locate and correct the hallucination-triggering video segments. Arbitrary or misaligned replacements cannot achieve comparable gains.

\subsection{Comparison with Alternatives}
\label{sec:comparison_alternatives}

Beyond analyzing feature replacement strategies, we further compare HALCON with alternative hallucination reduction methods that operate at different modality levels. For robustness, all
evaluations are conducted on the five Kling-Audio-Eval sublabels where the baseline model exhibits the highest hallucination rates. We report both the average IH@dur and the highest IH@dur observed within each setting, capturing overall hallucination tendency as well as worst case behavior.

\begin{figure}[h]
  \centering
  \includegraphics[width=1\linewidth]{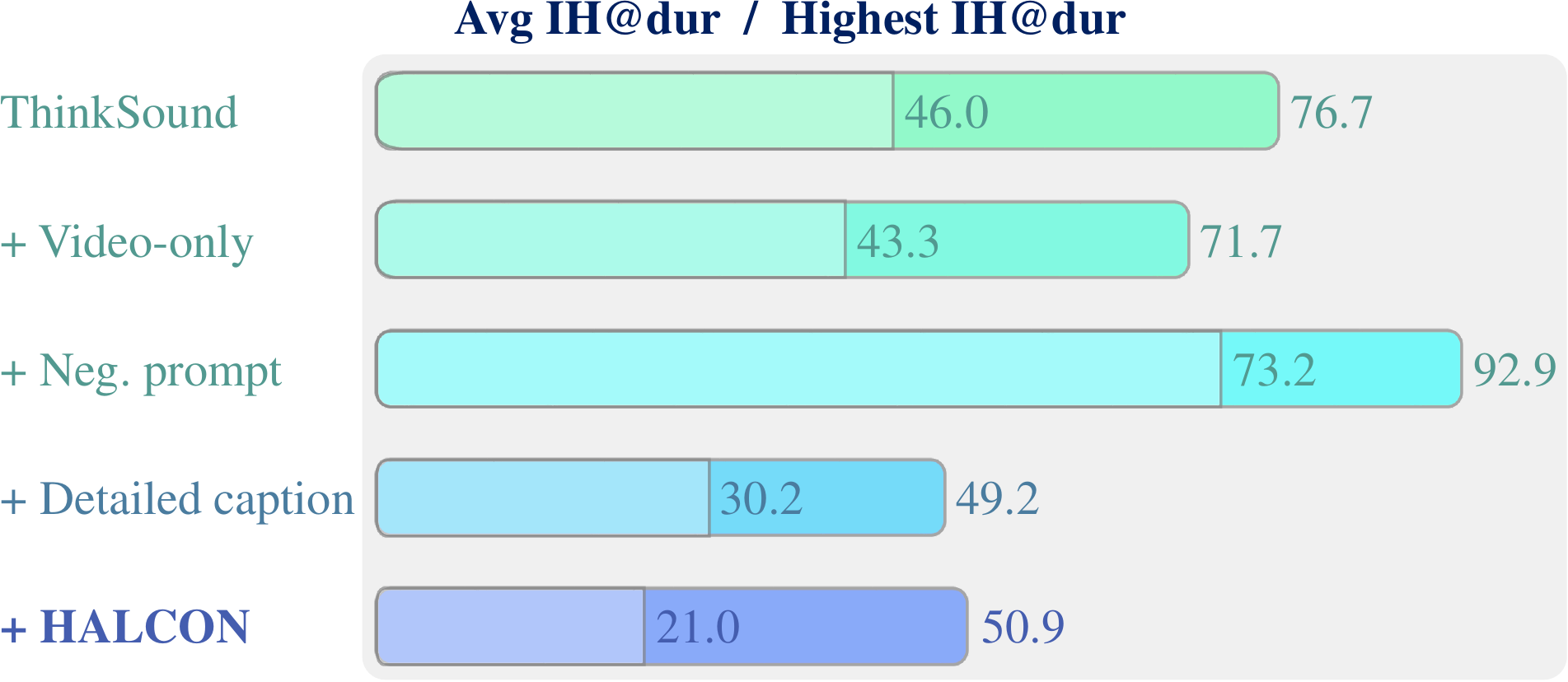}
  \caption{Comparison with alternative hallucination reduction methods. HALCON achieves the best IH@dur.}

  \label{fig:method_ablation}
\end{figure}

As shown in Figure~\ref{fig:method_ablation}, HALCON achieves the lowest average IH@dur and
substantially reduces the highest IH@dur among all evaluated methods. This demonstrates
that HALCON effectively suppresses hallucinations not only on average but also in extreme
failure cases, highlighting its robustness on the most challenging categories.

However, HALCON is not universally superior. In certain object-centric categories, detailed
captions achieve stronger reductions in hallucination, suggesting that fine-grained
textual descriptions provide semantic constraints that feature-level correction alone
cannot supply.

These complementary behaviors indicate that HALCON and text-driven approaches address
different aspects of hallucination. Combining input-level textual guidance (e.g., detailed
captions) with our posterior, feature-level correction (HALCON) may yield an even more
powerful and reliable method for suppressing hallucinations across diverse scenarios.

\section{Conclusion}

In this work, we investigate the problem of \emph{Insertion Hallucination} (IH) in V2A generation cases where models produce audio events without any visual support and demonstrate that this phenomenon is both common and systematically overlooked by existing evaluation metrics. To quantify IH, we introduce an end-to-end evaluation framework that combines a multi-detector ensemble with two metrics, IH@vid and IH@dur, enabling reliable measurement of both the prevalence and severity of hallucinations across datasets.

To mitigate IH, we propose HALCON, a three-stage inference-time conditioning method that identifies hallucinated segments, suppresses their misleading visual cues, and regenerates audio using corrected features. Extensive experiments show that HALCON consistently reduces hallucination rates by a large margin while maintaining and in some cases improving distributional, semantic, and synchronization quality.

Our findings highlight IH as a fundamental reliability challenge for V2A systems and demonstrate that addressing hallucinations requires going beyond conventional alignment oriented metrics. We hope this work encourages future research on building more trustworthy, visually grounded audio generation models and developing evaluation protocols that better reflect real-world deployment needs.





{
    \small
    \bibliographystyle{ieeenat_fullname}
    \bibliography{main}

@String(ECCV= {Eur. Conf. Comput. Vis.})

@String(ICASSP=	{ICASSP})

@String(ECCV  = {ECCV})

@inproceedings{vggsounder,
  title={VGGSounder: Audio-Visual Evaluations for Foundation Models},
  author={Zverev, Daniil and Wiedemer, Thadd{\"a}us and Prabhu, Ameya and Bethge, Matthias and Brendel, Wieland and Koepke, A},
  journal={arXiv preprint arXiv:2508.08237},
  year={2025}
}

@inproceedings{inaseg,
  author = {Doukhan, David and Lechapt, Eliott and Evrard, Marc and Carrive, Jean},
  title = {INA’S MIREX 2018 MUSIC AND SPEECH DETECTION SYSTEM},
  year = {2018},
  booktitle={Music Information Retrieval Evaluation eXchange (MIREX 2018)}
}

@misc{yamnet,
  author       = {Ellis, Dan and Hershey, Shawn and Weiss, Ron J. and Wilson, Kevin and Moore, J. R.},
  title        = {YAMNet: Yet Another MobileNet for Audio Event Classification},
  year         = {2019},
  publisher    = {GitHub},
  journal      = {TensorFlow Models - GitHub repository},
  howpublished = {\url{https://github.com/tensorflow/models/tree/master/research/audioset/yamnet}},
  commit       = {b362984}
}

@inproceedings{panns,
  title={Panns: Large-scale pretrained audio neural networks for audio pattern recognition},
  author={Kong, Qiuqiang and Cao, Yin and Iqbal, Turab and Wang, Yuxuan and Wang, Wenwu and Plumbley, Mark D},
  journal={IEEE/ACM Transactions on Audio, Speech, and Language Processing},
  volume={28},
  pages={2880--2894},
  year={2020},
  publisher={IEEE}
}

@InProceedings{vggsound,
  author       = "Honglie Chen and Weidi Xie and Andrea Vedaldi and Andrew Zisserman",
  title        = "VGGSound: A Large-scale Audio-Visual Dataset",
  booktitle    = "International Conference on Acoustics, Speech, and Signal Processing (ICASSP)",
  year         = "2020",
}

@article{klingfoley,
  title={Kling-Foley: Multimodal Diffusion Transformer for High-Quality Video-to-Audio Generation},
  author={Wang, Jun and Zeng, Xijuan and Qiang, Chunyu and Chen, Ruilong and Wang, Shiyao and Wang, Le and Zhou, Wangjing and Cai, Pengfei and Zhao, Jiahui and Li, Nan and others},
  journal={arXiv preprint arXiv:2506.19774},
  year={2025}
}

@inproceedings{fid,
  title        = {GANs trained by a two time-scale update rule converge to a local Nash equilibrium},
  author       = {Heusel, Martin and Ramsauer, Hubert and Unterthiner, Thomas and Nessler, Bernhard and Hochreiter, Sepp},
  booktitle    = {Advances in Neural Information Processing Systems (NeurIPS)},
  year         = {2017}
}

@inproceedings{vggish,
  title        = {CNN architectures for large-scale audio classification},
  author       = {Hershey, Shawn and Chaudhuri, Sourish and Ellis, Daniel PW and Gemmeke, Jort F and Jansen, Aren and Moore, R Channing and Plakal, Manoj and Platt, Devin and Slaney, Malcolm and Weiss, Ron J and others},
  booktitle    = {2017 IEEE International Conference on Acoustics, Speech and Signal Processing (ICASSP)},
  year         = {2017},
  organization = {IEEE}
}

@inproceedings{isc,
  title        = {Improved techniques for training GANs},
  author       = {Salimans, Tim and Goodfellow, Ian and Zaremba, Wojciech and Cheung, Vicki and Radford, Alec and Chen, Xi},
  booktitle    = {Advances in Neural Information Processing Systems (NeurIPS)},
  year         = {2016}
}

@inproceedings{desync,
  title     = {Cross-modal deep clustering},
  author    = {Ruder, Matthias and Lipping, Johannes and Pirsiavash, Hamed},
  booktitle = {Winter Conference on Applications of Computer Vision (WACV)},
  year      = {2020}
}

@inproceedings{chatterjee2020sound2sight,
  title={Sound2sight: Generating visual dynamics from sound and context},
  author={Chatterjee, Moitreya and Cherian, Anoop},
  booktitle={European Conference on Computer Vision},
  pages={701--719},
  year={2020},
  organization={Springer}
}

@article{pascual2017segan,
  title={SEGAN: Speech enhancement generative adversarial network},
  author={Pascual, Santiago and Bonafonte, Antonio and Serra, Joan},
  journal={arXiv preprint arXiv:1703.09452},
  year={2017}
}

@article{ferreira2022generation,
  title={On the generation of realistic synthetic petrographic datasets using a style-based GAN},
  author={Ferreira, Ivan and Ochoa, Luis and Koeshidayatullah, Ardiansyah},
  journal={Scientific Reports},
  volume={12},
  number={1},
  pages={12845},
  year={2022},
  publisher={Nature Publishing Group UK London}
}

@article{luo2023diff,
  title={Diff-foley: Synchronized video-to-audio synthesis with latent diffusion models},
  author={Luo, Simian and Yan, Chuanhao and Hu, Chenxu and Zhao, Hang},
  journal={Advances in Neural Information Processing Systems},
  volume={36},
  pages={48855--48876},
  year={2023}
}

@article{zhang2024foleycrafter,
  title={Foleycrafter: Bring silent videos to life with lifelike and synchronized sounds},
  author={Zhang, Yiming and Gu, Yicheng and Zeng, Yanhong and Xing, Zhening and Wang, Yuancheng and Wu, Zhizheng and Chen, Kai},
  journal={arXiv preprint arXiv:2407.01494},
  year={2024}
}

@inproceedings{cheng2025mmaudio,
  title={MMAudio: Taming Multimodal Joint Training for High-Quality Video-to-Audio Synthesis},
  author={Cheng, Ho Kei and Ishii, Masato and Hayakawa, Akio and Shibuya, Takashi and Schwing, Alexander and Mitsufuji, Yuki},
  booktitle={Proceedings of the Computer Vision and Pattern Recognition Conference},
  pages={28901--28911},
  year={2025}
}

@inproceedings{chen2025video,
  title={Video-guided foley sound generation with multimodal controls},
  author={Chen, Ziyang and Seetharaman, Prem and Russell, Bryan and Nieto, Oriol and Bourgin, David and Owens, Andrew and Salamon, Justin},
  booktitle={Proceedings of the Computer Vision and Pattern Recognition Conference},
  pages={18770--18781},
  year={2025}
}

@article{liu2025thinksound,
  title={ThinkSound: Chain-of-Thought Reasoning in Multimodal Large Language Models for Audio Generation and Editing},
  author={Liu, Huadai and Wang, Jialei and Luo, Kaicheng and Wang, Wen and Chen, Qian and Zhao, Zhou and Xue, Wei},
  journal={arXiv preprint arXiv:2506.21448},
  year={2025}
}

@article{lee2022factuality,
  title={Factuality enhanced language models for open-ended text generation},
  author={Lee, Nayeon and Ping, Wei and Xu, Peng and Patwary, Mostofa and Fung, Pascale N and Shoeybi, Mohammad and Catanzaro, Bryan},
  journal={Advances in Neural Information Processing Systems},
  volume={35},
  pages={34586--34599},
  year={2022}
}

@article{mundler2023self,
  title={Self-contradictory hallucinations of large language models: Evaluation, detection and mitigation},
  author={M{\"u}ndler, Niels and He, Jingxuan and Jenko, Slobodan and Vechev, Martin},
  journal={arXiv preprint arXiv:2305.15852},
  year={2023}
}

@inproceedings{Li-hallucination-2023,
  title={Evaluating Object Hallucination in Large Vision-Language Models},
  author={Li, Yifan and Du, Yifan and Zhou, Kun and Wang, Jinpeng and Zhao, Wayne Xin and Wen, Ji-Rong},
  booktitle={The 2023 Conference on Empirical Methods in Natural Language Processing},
  year={2023},
  url={https://openreview.net/forum?id=xozJw0kZXF}
}

@article{chen2024multi,
  title={Multi-object hallucination in vision language models},
  author={Chen, Xuweiyi and Ma, Ziqiao and Zhang, Xuejun and Xu, Sihan and Qian, Shengyi and Yang, Jianing and Fouhey, David and Chai, Joyce},
  journal={Advances in Neural Information Processing Systems},
  volume={37},
  pages={44393--44418},
  year={2024}
}

@article{compagnoni2025mitigating,
  title={Mitigating Hallucinations in Multimodal LLMs via Object-aware Preference Optimization},
  author={Compagnoni, Alberto and Caffagni, Davide and Moratelli, Nicholas and Baraldi, Lorenzo and Cornia, Marcella and Cucchiara, Rita},
  journal={arXiv preprint arXiv:2508.20181},
  year={2025}
}

@article{sung2024avhbench,
  title={Avhbench: A cross-modal hallucination benchmark for audio-visual large language models},
  author={Sung-Bin, Kim and Hyun-Bin, Oh and Lee, JungMok and Senocak, Arda and Chung, Joon Son and Oh, Tae-Hyun},
  journal={arXiv preprint arXiv:2410.18325},
  year={2024}
}

@inproceedings{kushwaha2025vintage,
  title={Vintage: Joint video and text conditioning for holistic audio generation},
  author={Kushwaha, Saksham Singh and Tian, Yapeng},
  booktitle={Proceedings of the Computer Vision and Pattern Recognition Conference},
  pages={13529--13539},
  year={2025}
}

@inproceedings{chen2024action2sound,
  title={Action2sound: Ambient-aware generation of action sounds from egocentric videos},
  author={Chen, Changan and Peng, Puyuan and Baid, Ami and Xue, Zihui and Hsu, Wei-Ning and Harwath, David and Grauman, Kristen},
  booktitle={European Conference on Computer Vision},
  pages={277--295},
  year={2024},
  organization={Springer}
}

@article{zverev2025vggsounder,
  title={VGGSounder: Audio-Visual Evaluations for Foundation Models},
  author={Zverev, Daniil and Wiedemer, Thadd{\"a}us and Prabhu, Ameya and Bethge, Matthias and Brendel, Wieland and Koepke, A},
  journal={arXiv preprint arXiv:2508.08237},
  year={2025}
}

@article{madaan2023self,
  title={Self-refine: Iterative refinement with self-feedback},
  author={Madaan, Aman and Tandon, Niket and Gupta, Prakhar and Hallinan, Skyler and Gao, Luyu and Wiegreffe, Sarah and Alon, Uri and Dziri, Nouha and Prabhumoye, Shrimai and Yang, Yiming and others},
  journal={Advances in Neural Information Processing Systems},
  volume={36},
  pages={46534--46594},
  year={2023}
}

@article{shinn2023reflexion,
  title={Reflexion: Language agents with verbal reinforcement learning, 2023},
  author={Shinn, Noah and Cassano, Federico and Labash, Beck and Gopinath, Ashwin and Narasimhan, Karthik and Yao, Shunyu},
  journal={URL https://arxiv. org/abs/2303.11366},
  volume={1},
  year={2023}
}

@article{huang2023large,
  title={Large language models cannot self-correct reasoning yet, 2024},
  author={Huang, Jie and Chen, Xinyun and Mishra, Swaroop and Zheng, Huaixiu Steven and Yu, Adams Wei and Song, Xinying and Zhou, Denny},
  journal={arXiv preprint arXiv:2310.01798},
  year={2023}
}

@InProceedings{tian2018ave,
  author={Tian, Yapeng and Shi, Jing and Li, Bochen and Duan, Zhiyao and Xu, Chenliang},
  title={Audio-Visual Event Localization in Unconstrained Videos},
  booktitle = {ECCV},
  year = {2018}
}
}

\newpage
\onecolumn
\section{Appendix}

\subsection{Hallucination Detection Pipeline}
\label{sec:appendix_detector}
\paragraph{Overview}
Our hallucination detection pipeline operates on fixed-length
clips of $9.98$\,s. Given an audio file, each detector
produces a set of segments
$\{(l,s,e)\}$ where $l \in \{\texttt{speech},\texttt{music}\}$
and $0 \le s < e \le T$.
For each clip we compute
\[
\text{IH@vid} = \mathbb{1}[d_{\mathrm{IH}} > 0], \qquad
\text{IH@dur} = d_{\mathrm{IH}} / T,
\]
where $d_{\mathrm{IH}}$ is the union duration of all
speech/music segments and $T=9.98$\,s is the clip length.
Union duration is computed after merging overlapping
speech/music intervals.

For robustness we use three off-the-shelf detectors:
inaSpeechSegmenter, YAMNet and PANNs-CNN14. All detectors
are run in CPU-only mode with restricted threading to avoid
contention in multi-processing.
For each detector we keep only speech/music segments and
store them in CSV/JSON files for later fusion and analysis.
Unless otherwise specified, we use the default weights
released by the corresponding authors.

\paragraph{inaSpeechSegmenter (ISS)}
We use the official \texttt{inaSpeechSegmenter} implementation
with the following settings:
\begin{itemize}
  \item \textbf{Engine:} \texttt{vad\_engine='smn'}, \texttt{detect\_gender=False}.
  \item \textbf{Classes:} From the raw ISS output we keep only
        segments whose label is either \texttt{speech} or \texttt{music}.
  \item \textbf{Post-processing:} We round start/end times to milliseconds
        and discard non-positive-length segments.
\end{itemize}
ISS directly outputs contiguous segments, so we do not change
its internal thresholds; all later temporal smoothing is performed
by our shared post-processing described below.

\paragraph{YAMNet}
For YAMNet we use the TensorFlow Hub model
\texttt{google/yamnet/1} and compute frame-level scores at
16\,kHz. Each frame corresponds to a 0.96\,s receptive field
with 50\% overlap; we approximate the frame hop as
$t_{\mathrm{hop}} = 0.48$\,s.

We map the original AudioSet labels to speech and music
using curated label lists. Let $\mathcal{I}_{\mathrm{sp}}$ and
$\mathcal{I}_{\mathrm{mu}}$ denote the indices of classes
belonging to speech and music respectively. For each
frame $i$ we compute
\[
p^{(i)}_{\mathrm{sp}} = \max_{c \in \mathcal{I}_{\mathrm{sp}}} p^{(i)}_c,
\qquad
p^{(i)}_{\mathrm{mu}} = \max_{c \in \mathcal{I}_{\mathrm{mu}}} p^{(i)}_c.
\]
A frame is marked as speech if
$p^{(i)}_{\mathrm{sp}} \ge 0.40$ and as music if
$p^{(i)}_{\mathrm{mu}} \ge 0.30$.
Each positive frame $i$ is converted to a tentative segment
$[s_i, e_i] = [i \cdot t_{\mathrm{hop}}, (i+1)\cdot t_{\mathrm{hop}}]$
with the corresponding label.
These frame-level segments are then merged using our
generic merging function (described below) with
$\texttt{min\_dur}=0.20$\,s and $\texttt{min\_gap}=0.15$\,s.

\paragraph{PANNs (CNN14)}
We use the \texttt{SoundEventDetection} interface from
\texttt{panns\_inference} with the CNN14 backbone pretrained
on AudioSet. Audio is resampled to 32\,kHz and fed as
a single-channel waveform.

The model outputs a frame-wise tensor of shape
$T \times C$. We normalise label names to lower case and
map them into the same speech and music groups as in the
main paper. Let $\mathcal{J}_{\mathrm{sp}}$ and
$\mathcal{J}_{\mathrm{mu}}$ denote the corresponding index sets.
For each time frame $i$ we compute
\[
q^{(i)}_{\mathrm{sp}} = \max_{c \in \mathcal{J}_{\mathrm{sp}}} q^{(i)}_c,
\qquad
q^{(i)}_{\mathrm{mu}} = \max_{c \in \mathcal{J}_{\mathrm{mu}}} q^{(i)}_c.
\]
We mark frame $i$ as speech if $q^{(i)}_{\mathrm{sp}} \ge 0.40$
and as music if $q^{(i)}_{\mathrm{mu}} \ge 0.30$.
Let $T$ be the number of frames and $L$ the clip duration;
the corresponding frame hop is $t_{\mathrm{hop}} = L/T$.
As with YAMNet, positive frames are converted into segments
$[s_i, e_i]$ and merged by the common post-processing
with $\texttt{min\_dur}=0.20$\,s and $\texttt{min\_gap}=0.15$\,s.

\paragraph{Temporal Smoothing and Merging}
All detectors share the same temporal smoothing procedure.
Given a list of frame-level segments
$(s,e,l)$, we first sort them by start time
and then iteratively merge consecutive segments with the
same label if the gap between them is shorter than
$\texttt{min\_gap}=0.15$\,s. Segments shorter than
$\texttt{min\_dur}=0.20$\,s are discarded.
This removes spurious short bursts and enforces temporal
consistency across detectors.

\paragraph{Multi-detector Fusion}
To combine the three detectors, we discretise the time
axis into a uniform grid with step $\Delta t = 0.02$\,s.
For each detector, we project its speech/music segments
onto this grid (with a tolerance of 0.05\,s) and obtain
binary vote matrices of shape $K \times N$, where
$K$ is the number of detectors and $N$ is the number
of grid frames.

We consider three fusion rules:
\begin{itemize}
  \item \textbf{OR}: a frame is positive if at least one detector votes positive;
  \item \textbf{AND}: a frame is positive only if all detectors agree;
  \item \textbf{MV (Majority Vote)}: a frame is positive if at least
        $\lceil K/2 \rceil$ detectors vote positive.
\end{itemize}
Positive frames are then converted back to continuous
segments and merged with the same
$\texttt{min\_dur}=0.20$\,s and $\texttt{min\_gap}=0.15$\,s
constraints. All main results in the paper use the
\textbf{MV} fusion rule, which we show in Sec.~5.2 to offer
the best precision--recall trade-off on our human-annotated
validation set.

\subsection{Human Annotation Pipeline and Use for Metric Validation}
\label{sec:appendix_human}

\paragraph{Sampling.}
We construct a human-annotated validation set from the outputs of two representative V2A systems, MMAudio and ThinkSound, on the Kling-Audio-Eval benchmark.
For each sublabel in Kling-Audio-Eval, we randomly sample $20$ generated audio clips per model, yielding over $900$ clips in total and more than $9{,}000$ seconds of audio.\footnote{Counts refer to post-filtered items that pass loading and playback checks.}
This per-sublabel sampling strategy provides broad coverage of object- and scene-centric categories while keeping the annotation workload tractable.

\begin{figure}[t]
  \centering
  \includegraphics[width=0.92\linewidth]{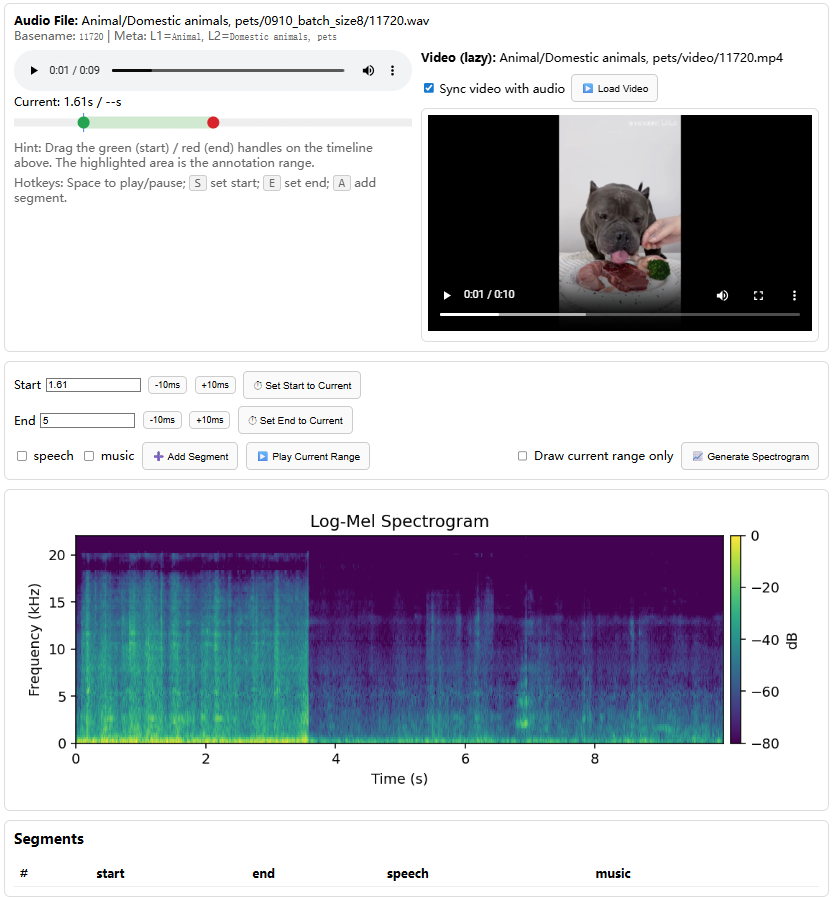}
  \caption{Web interface for human annotation. Annotators can view the paired video and the spectrogram, select arbitrary time spans for playback, and mark multiple speech or music segments with $0.01$\,s precision.}
  \label{fig:annotation_ui}
\end{figure}

\paragraph{Annotation Interface.}
Annotators use a web-based tool that supports precise temporal labeling (Figure~\ref{fig:annotation_ui}).
They can view the paired video, inspect the mel-spectrogram, scrub or play arbitrary time spans, and add multiple speech or music segments per clip.
Timestamps are recorded at $0.01$\,s resolution, and the interface enforces basic consistency constraints.

\paragraph{Labeling Guidelines.}
Annotators mark a segment only if the generated audio contains audible speech or music without a plausible visual source.
Ambiguous non-speech vocalisations are ignored unless clearly constituting speech or singing.
Environmental noise and percussive impacts are not labeled.
Segments shorter than $0.20$\,s are discouraged, and gaps shorter than $0.15$\,s between adjacent segments of the same type may be merged.

\paragraph{Boundary Conventions.}
Boundaries align to perceptual onset/offset at $0.01$\,s granularity.
Adjacent labels separated by gaps $<0.15$\,s are merged.
During evaluation of detector accuracy, we apply a symmetric tolerance of $0.05$\,s around annotated boundaries (Section~\ref{sec:appendix_validation}).

\paragraph{Annotator Roles.}
Two primary annotators independently label every clip, and a third adjudicator resolves disagreements.
All annotators work with over-ear headphones in a quiet room, performing a brief warm-up on five clips with discussion of edge cases.

\paragraph{Quality Control.}
Checks include schema validation (types, ordering), consistency validation (no impossible overlaps), and cross-annotator comparison (IoU and clip-level agreement).
If disagreement persists, the adjudicator produces the final consensus annotation.

\paragraph{Aggregation and Export.}
\label{sec:appendix_annotation:aggregation}
We export per-clip annotations with fields: clip\_id, model, sublabel, segment\_type, start, end.
Timestamps are rounded to $0.01$\,s.
Post-processing merges gaps $<0.15$\,s and removes fragments shorter than $0.20$\,s unless manually retained.
All annotations remain in the clip’s native time base.

\paragraph{Use in Validating IH Metrics.}
\label{sec:appendix_validation}
For metric validation, we treat hallucination detection as a framewise binary classification problem on a fixed temporal grid.
Both human annotations and detector outputs are converted into binary masks at a resolution of $0.01$\,s over a 10\,s clip.
Adjudicated human segments are first clamped to $[0,10]$ and rasterized into a ground-truth mask $m^{\ast}$.
For each detector (ina, panns, yamnet), we load its CSV file, select all rows associated with the current clip, clamp predicted timestamps, and rasterize them into a mask $m$ on the same grid.
An optional tolerance $\tau$ can be applied via one-dimensional binary dilation, though the main results use $\tau=0$.

For each clip and each method (three base detectors and three fusion strategies: AND, OR, and majority vote), we compute per-frame true positives, false positives, and false negatives as
$\mathrm{TP}=\sum (m^{\ast} \wedge m)$,
$\mathrm{FP}=\sum (\neg m^{\ast} \wedge m)$,
$\mathrm{FN}=\sum (m^{\ast} \wedge \neg m)$,
and accumulate these counts over all annotated clips.
We then compute precision, recall, and IoU, and evaluate the $F_{\beta}$ score over $\beta \in [0,1]$:
\[
F_{\beta} = (1+\beta^{2}) \cdot 
\frac{\mathrm{Precision} \cdot \mathrm{Recall}}
{\beta^{2}\,\mathrm{Precision} + \mathrm{Recall}}.
\]
Varying $\beta$ allows us to smoothly adjust the relative importance of precision and recall, enabling a more complete characterization of detector behavior.
This evaluation pipeline ensures strict temporal alignment between human judgments and detector predictions and provides a consistent basis for comparing individual detectors and fusion strategies.

\paragraph{Statistics.}
The final annotated set contains over $1{,}000$ clips (more than $10{,}000$ seconds).
Speech and music prevalence varies substantially across sublabels.
We report hallucination prevalence and average hallucinated duration in Section~\ref{sec:validation}.

\paragraph{Intended Use.}
This annotated set is used solely for evaluating the IH detection pipeline and is never used to train or tune HALCON.
All quantitative results relying on human labels reference the adjudicated split.

\subsection{Case Studies of Hallucinated Speech and Music}
\label{sec:appendix_case_study_all}

\begin{figure}[t]
    \centering
    \includegraphics[width=0.72\linewidth]{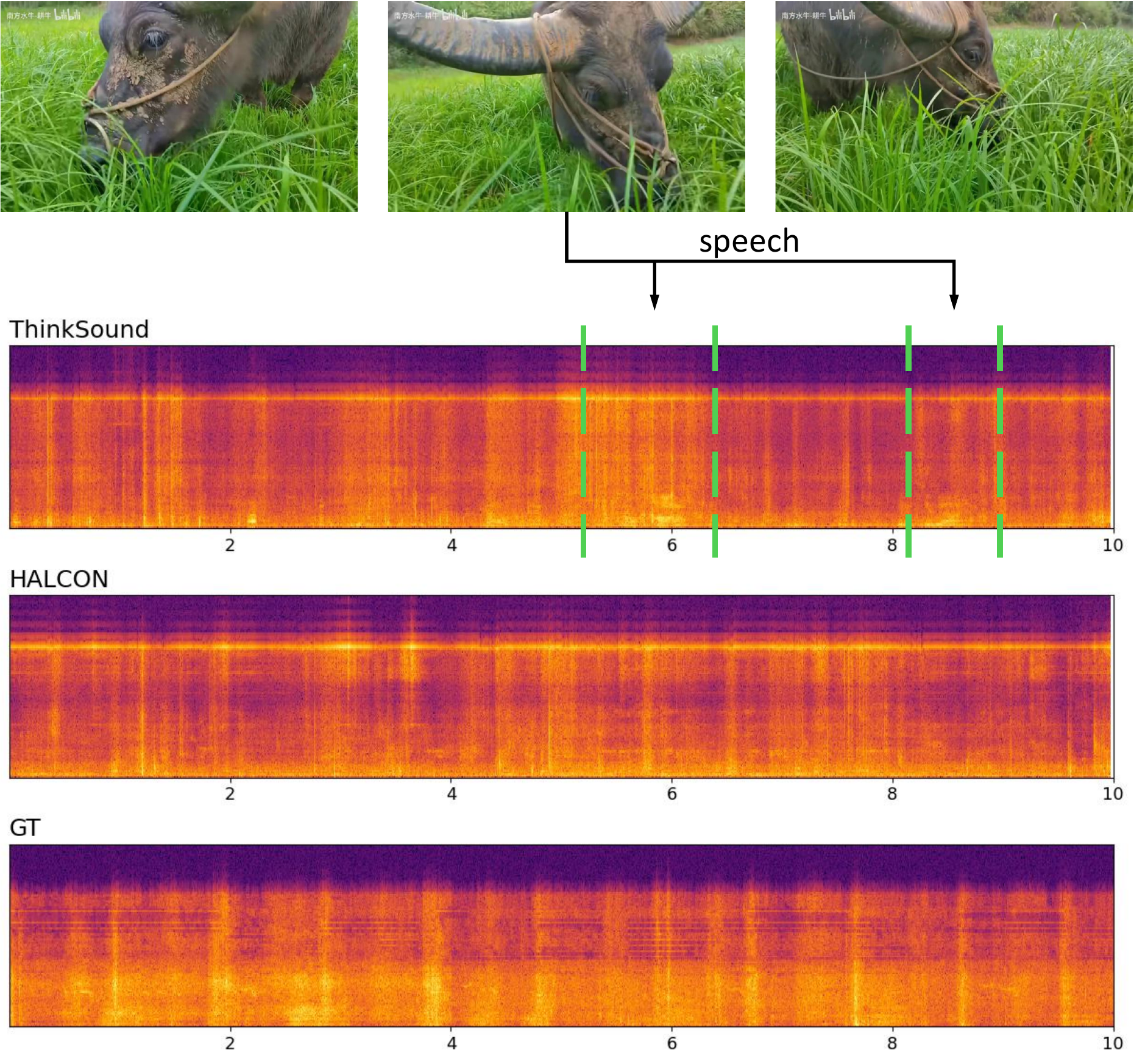}
    \caption{
    \textbf{Case study showing hallucinated speech in ThinkSound.}
    The video shows a buffalo grazing, a scene with no plausible speech source.
    ThinkSound introduces two spurious speech segments (indicated by green markers),
    while HALCON eliminates both and aligns closely with the ground-truth (GT) acoustic structure.
    }
    \label{fig:case_study_1}
\end{figure}

\paragraph{Case 1: Speech hallucination.}
Figure~\ref{fig:case_study_1} shows a representative example in which the video depicts a buffalo grazing in a field, a scene that contains no plausible speech source. Nevertheless, the audio produced by ThinkSound includes two hallucinated speech segments, whereas HALCON successfully suppresses both artifacts and remains consistent with the natural acoustic scene.

In the ThinkSound spectrogram, two speech-like regions emerge around \textbf{5.5--6.5\,s} and \textbf{8.2--9.0\,s}.  
The first region is faint: it contains several short, weak, and discontinuous high-frequency streaks that create only a vague vocal impression.  
The second region displays clear speech structure, including:
(i) stable formant-like trajectories between 300--3{,}000\,Hz,
(ii) periodic temporal patterns resembling syllabic rhythm, and
(iii) transient energy bursts indicative of consonant onsets.  
Listening confirms that the second segment corresponds to a clearly audible male voice saying ``\emph{fifty nine},'' which is unrelated to the visual scene and strongly disrupts perceptual coherence.

In contrast, HALCON exhibits no formant patterns, no periodic vocal structures, and no consonant-like transients throughout the clip.  
Its time--frequency texture remains stable and consistent with natural grazing sounds such as foliage rustling, low-frequency animal movement, and ambient outdoor noise.  
This matches the ground-truth (GT) spectrogram, which contains no speech activity.

\begin{figure}[t]
    \centering
    \includegraphics[width=0.72\linewidth]{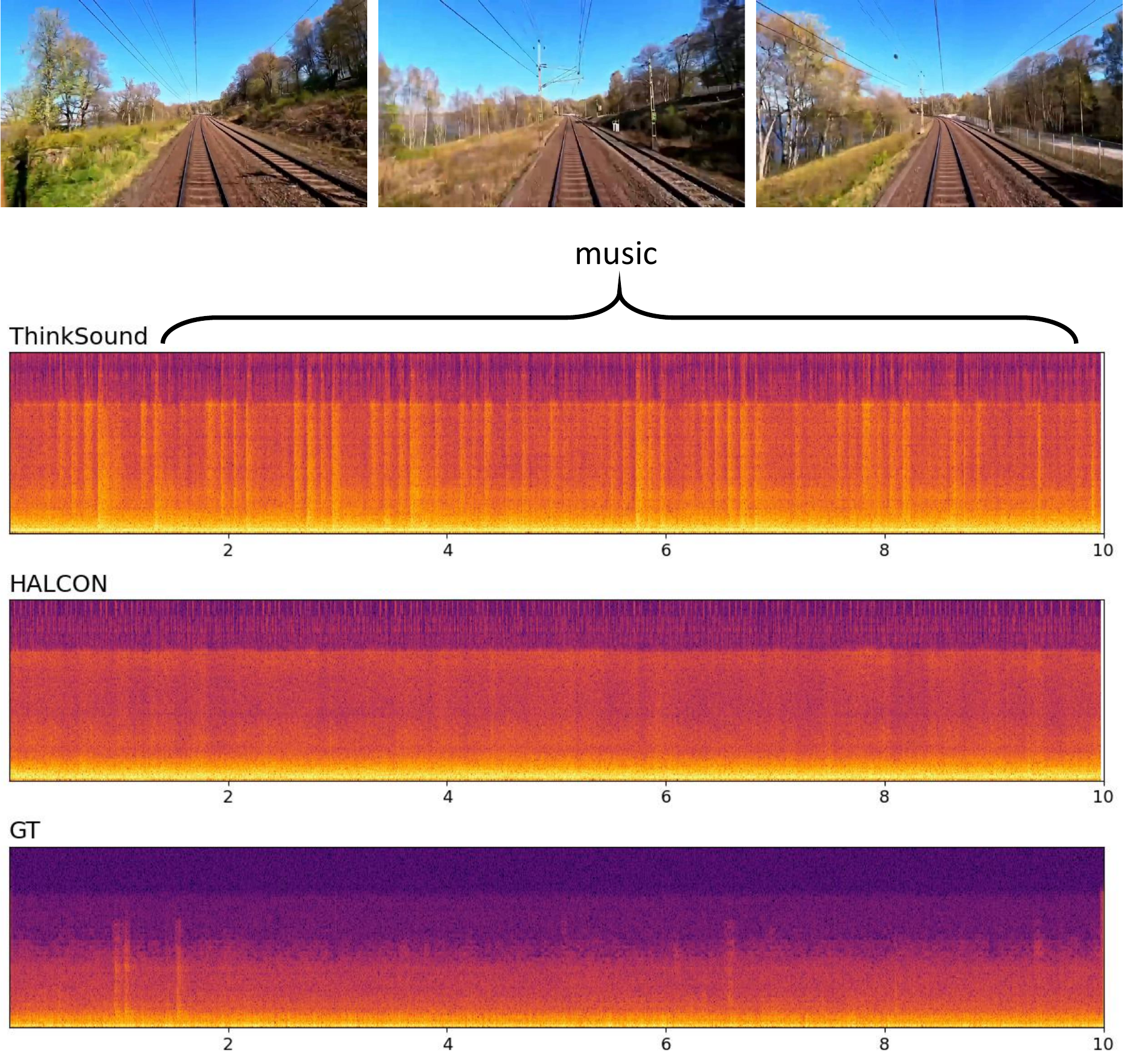}
    \caption{
    \textbf{Case study showing hallucinated music in ThinkSound.}
    The video records a train moving through a rural landscape, a scene with no plausible music source.
    ThinkSound introduces a long-duration music-like pattern that spans almost the entire clip,
    whereas HALCON suppresses these artifacts and produces an output that is consistent with the ground-truth (GT) ambient acoustic structure.
    }
    \label{fig:case_study_2}
\end{figure}

\paragraph{Case 2: Long-duration music hallucination.}
Figure~\ref{fig:case_study_2} provides another example, where the video shows a train moving through a rural landscape.  
The true audio in such a scene is dominated by low-frequency rolling noise, wind, and occasional broadband transients, with no plausible source of music.  
However, ThinkSound generates a persistent music-like pattern that spans nearly the entire \textbf{0--10\,s} duration.

The hallucinated music is characterized by:
(i) dense vertical striations that resemble synthetic percussive or rhythmic elements,  
(ii) quasi-periodic temporal repetition suggestive of a musical beat or harmonic cycle, and  
(iii) stable high-frequency energy bands inconsistent with environmental noise.  
These patterns collectively form a clearly perceptible ``background music'' layer that overrides the natural ambience expected from the scene.

HALCON suppresses these artifacts entirely.  
Its spectrogram exhibits smooth, broadband low-frequency energy consistent with train motion, sparse and irregular high-frequency components typical of wind and environmental texture, and no periodic or harmonic spectral structures.  
Compared with the GT spectrogram, HALCON closely preserves the authentic acoustic characteristics of the scene and avoids introducing any music-like artifacts.

\paragraph{Summary.}
Across both speech and music examples, ThinkSound produces salient, perceptually dominant hallucinated content—either short but distinct speech or long-duration music.  
HALCON consistently removes such spurious vocal and musical artifacts while maintaining scene-appropriate acoustic structure, achieving outputs that are both perceptually and physically aligned with the visual input.

\end{document}